\documentclass[fleqn,usenatbib]{mnras}

\DeclareRobustCommand{\VAN}[3]{#2}
\let\VANthebibliography\thebibliography
\def\thebibliography{\DeclareRobustCommand{\VAN}[3]{##3}\VANthebibliography}

\usepackage{graphicx}	

\title[Leptocline as a Shallow Substructure of NSSL]{Leptocline as a Shallow Substructure of Near-Surface Shear Layer in 3D Radiative Hydrodynamic Simulations}

\author[I. N. Kitiashvili et al.]{
	Irina N. Kitiashvili,$^{1}$\thanks{E-mail: Irina.N.Kitiashvili@nasa.gov}
    A. G. Kosovichev$^2$,
    A. A. Wray$^1$,
    V. M. Sadykov$^3$, 
    G. Guerrero$^{2,4}$
	\\
$^{1}$NASA Ames Research Center, Moffett Field, MS 258-5, Mountain View, 94035, USA\\
$^{2}$New Jersey Institute of Technology, Newark, NJ 07102, USA\\
$^{3}$Georgia State University, Atlanta, 30303 GA, USA\\
$^{4}$Universidade Federal de Minas Gerais, Belo Horizonte, MG 31270, Brazil}

\date{Accepted XXX. Received YYY; in original form ZZZ}

\pubyear{2022}

\begin{document}
\label{firstpage}
\maketitle

\begin{abstract}
 Understanding effects driven by rotation in the solar convection zone is essential for many problems related to solar activity, such as the formation of differential rotation, meridional circulation, and others. We analyze realistic 3D radiative hydrodynamics simulations of solar subsurface dynamics in the presence of rotation in a local domain 80 Mm wide and 25 Mm deep, located at 30 degrees latitude. The simulation results reveal the development of a shallow 10-Mm deep substructure of the Near-Surface Shear Layer (NSSL), characterized by a strong radial rotational gradient and self-organized meridional flows. This shallow layer (``leptocline") is located in the hydrogen ionization zone associated with enhanced anisotropic overshooting-type flows into a less unstable layer between the H and HeII ionization zones. We discuss current observational evidence of the presence of the leptocline and show that the radial variations of the differential rotation and meridional flow profiles obtained from the simulations  in this layer qualitatively agree with helioseismic observations.
\end{abstract}




\section{Introduction}
The discovery of solar rotation by Galileo and its variation with latitude by Christoph Scheiner through tracking sunspots across the solar disk was the first indication of complex processes associated with the Sun's interior dynamics and activity. 
Intensive studies of global dynamics using spectroscopic and sunspot observations and derivation of the properties of differential rotation stimulated theoretical investigations and data analysis developments.
Variations of the solar rotation in the radial direction have been anticipated from early surface observations of magnetic tracers, active regions, and supergranules \citep[e.g.,][]{Wilcox1970,Howard1970,Howard1975,Foukal1976}. The initial inferences of the solar subsurface rotation by \cite{Deubner1979} revealed the existence of a shallow subsurface layer with a sharp increase of the rotation rate with depth.
Further development of helioseismology techniques to probe solar internal structure and dynamics made it possible to investigate the properties and evolution of solar differential rotation, meridional flow, and Near-Surface Shear Layer (NSSL), occupying approximately the upper 15-20\% of the convection zone  \citep[e.g.,][]{Gough1981,Kosovichev1997,Kosovichev2006,Thompson1996,Thompson2003,Howe2009,Zhao2013, Basu2019}. For the history of the discovery and studies of the NSSL, we refer the readers to the review article by \citet{Howe2009}.
Helioseismic observations reveal that the near-surface shear layer extends from the equator to high latitudes, and that the mean radial gradient of the angular velocity is almost constant, $|d\log\Omega/d\log r| \approx 1$ \citep{Corbard2002,Barekat2014}. However, recent global helioseismology results show that the gradient of rotation is not uniform but increases closer to the surface \citep{Reiter2020,Antia2022}. In addition, local helioseismology has revealed variations of the meridional circulation with depth in the NSSL \citep{Zhao2004,Komm2018}. 

Reproducibility of the observed global solar flows in simulations remains a challenging task and topic of hot debate. For instance, \cite{Miesch2011} suggested that the distinct NSSL is maintained by a transition from baroclinic to turbulent stresses. 
Traditionally, the effects of solar rotation are modeled on global scales using an anelastic approximation that allows simulating the whole spherical convection zone, but these models exclude the near-surface layers where the anelastic approximation is not valid \citep[e.g.,][]{Gilman1979,Brun2002,Brun2011}. While the anelastic models of \citet{Guerrero2016a,Guerrero2019} and \citet{Stejko2020} were able to reproduce a solar-like NSSL in 3D models, other global simulations reproduced it only in the near-equatorial region \citep[e.g.,][]{Matilsky2019,Hotta2022}, contrary to the solar observations. 
A likely reason is that global-Sun simulations are not yet capable of resolving the essential dynamical scales in the NSSL, particularly close to the solar surface.
To address this issue, \citet{Barekat2021} performed simulations for several near-surface local patches located at different latitudes, using a simplified force-driven turbulence model and neglecting density stratification and energy transport. They showed the importance of local Reynolds stresses in the formation of the rotational shear and meridional circulation. However, these simulations also reproduced the rotational shear only in the equatorial region.

To gain insight into the structure and dynamics of the upper layers of the convection zone in the presence of rotation, we perform local 3D radiative hydrodynamic modeling of the uppermost layers of solar convection. The computational model includes effects of compressibility, radiative energy transport, and subgrid-scale turbulence and reproduces solar convection with a high degree of realism. Currently, such simulations require substantial computational resources and cannot be performed for the whole spherical Sun. Therefore, the computational domain is limited to a local region, in this case, a rectangular volume 80 Mm wide and 25 Mm deep, located at 30 degrees latitude. 

For the first time that we are aware, such a realistic simulation of rotational effects in solar subsurface convection produces the formation of rotational shear and meridional circulation at mid-latitudes, in agreement with observations. Furthermore, they show that the structure of the NSSL is not uniform but contains a sharp shear layer in the top $\sim 8$~Mm, which we identify as a `leptocline', a previously conjectured shallow layer beneath the solar surface \citep{Godier2001,Rozelot2009b}. We show that the simulation results are in good qualitative agreement with helioseismic inferences and predict new features in variations of the meridional circulation and other properties with depth that can be tested by helioseismology.

The presentation of the results begins with a brief description of the numerical setup (Section~\ref{setup}). Then, Section~\ref{termdyn} describes the thermodynamic and dynamical properties of turbulent and mean flows, particularly the self-formed subsurface {\bf rotational} shear layer and meridional flows. Finally, Section~\ref{conclusion} discusses the main findings and compares them with observations.

\section{Computational Setup}\label{setup}
We perform 3D radiative hydrodynamic simulations using the StellarBox code \citep{Wray2018}. The formulation of the StellarBox code includes the fully compressible MHD equations from first principles plus radiative transfer. Rotational effects are modeled in the $f$-plane approximation. The computational model takes into account the realistic chemical composition and equation of state and uses a large-eddy simulation (LES) treatment of subgrid turbulent transport. Subgrid turbulence models \citep{Smagorinsky1963,Moin1991} are critical for accurately describing small-scale energy dissipation and transport. The radiative transfer calculations are performed for four spectral bins; ray-tracing along 18 directional rays \citep{Feautrier1964} is implemented using the long-characteristics method. The wavelength-dependent opacity code and data are provided by the Opacity Project \citep{Seaton1995b,Badnell2005}. The simulations are performed in Cartesian geometry, and the lateral boundary conditions of the computational domain are periodic. The top boundary condition is implemented using a characteristic method \citep[e.g.,][]{Sun1995}; mass, momentum, and energy are allowed to pass through this boundary in either direction as determined by the characteristic decomposition. Inward radiative flux at the top boundary is taken to be zero. The bottom boundary of the computational domain is closed to mass and momentum flux through the imposition of a zero normal velocity in the bottom plane of the domain. The energy input from the interior of the Sun is imposed as a steady and horizontally uniform energy flux across the bottom plane.  In addition, total mass and momentum conservation are imposed by introducing uniform mass and momentum fluxes at the bottom boundary to compensate for any gain or loss of these quantities through the top boundary.  \emph{Convective} energy loss through the top boundary is similarly compensated. \emph{Total} energy conservation is not imposed in this manner but is allowed to reach a statistical balance between the constant imposed energy input at the bottom and the calculated time-varying radiative loss through the top.  The result is heat transfer through the domain by a mixture of convective (generally turbulent) motion and radiative transport. The simulations are initialized as a 3-D perturbation of a hydrostatically balanced vertical profile from a standard solar model of the interior structure and the lower atmosphere \citep{Christensen-Dalsgaard1996}.  The perturbations consist of spatially and directionally random $\sim 10$ cm/sec velocities.   The code details, implementation and testing are described by \citet{Wray2015,Wray2018}.

\begin{figure}
	\begin{center}
		\includegraphics[scale=0.7]{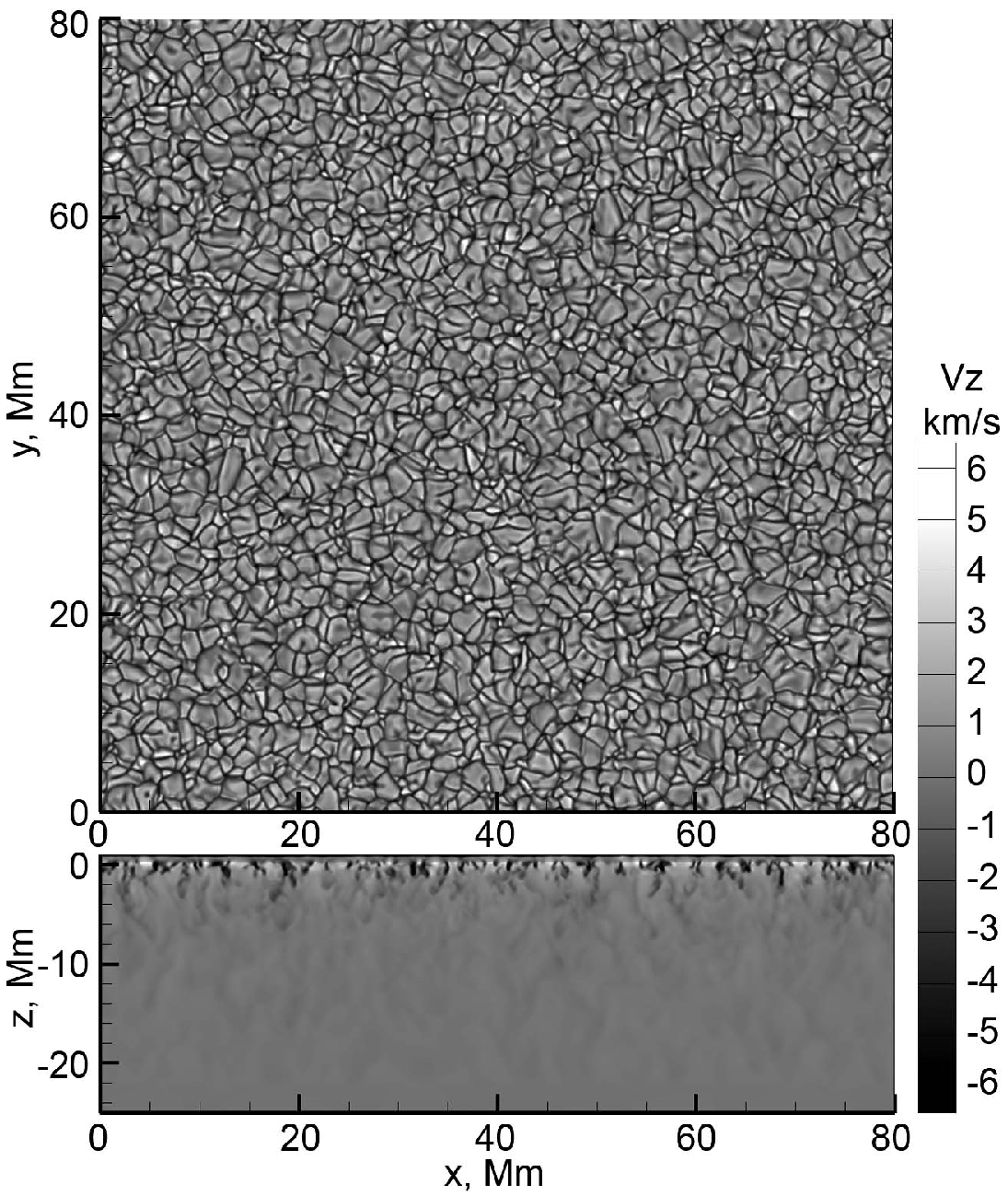}
	\end{center}
	\caption{Snapshot of the vertical velocity at the solar photosphere (top panel) and a vertical slice through the computational domain (bottom). \label{fig:vz}}
\end{figure}

In this paper, we present an analysis of a model with an imposed rotation corresponding to 30 degrees latitude. The horizontal size of the computational domain is 80 Mm $\times$ 80 Mm, and the vertical domain extends to a depth of 25 Mm (Fig.~\ref{fig:vz}).  The grid resolution is 100 km in the horizontal directions; the vertical resolution varies from 50 km in the photosphere and low atmosphere to 82 km near the bottom boundary. The computational $x$-axis is oriented in the azimuthal direction, and the $y$-axis is directed toward the North pole. The bottom 5 Mm of the computational domain were excluded from the analysis to avoid potential boundary-related effects. The model includes a 1 Mm high atmospheric layer. The extended duration of the simulation, over 250 hours, allows us to reach dynamically stationary conditions before investigating the influence of rotational effects. 
It corresponds to about 100 convective turnover times in the upper 10 Mm deep layer and about 40 turnover times in the 20 Mm deep layer, which is sufficient for reaching the stationary conditions in a broad range of an effective Coriolis number \citep{Barekat2021}.
The angular momentum and other parameters reached quasi-stationary states, and angular momentum variations in the last $100$ hours of the simulation run never exceeded 5\% (Fig.~\ref{fig:rossby}d).  For analysis, we use the last 24 hours of the computations. The data cubes are collected with a cadence of 45sec. 

\section{Properties of the Solar Convection in the Near-Surface Shear Layer}\label{termdyn}
Snapshots of the vertical velocity on the surface and in a radial slice of the computational domain are shown in Figure~\ref{fig:vz}. Detailed properties of the granulation structure and dynamics are discussed in our previous papers \citep[e.g.,][]{Kitiashvili2011b,Kitiashvili2013a,Kitiashvili2013,Kitiashvili2015}. In this paper, we focus on the effects of solar rotation.  

\begin{figure}
	\begin{center}
		\includegraphics[width=0.9\textwidth]{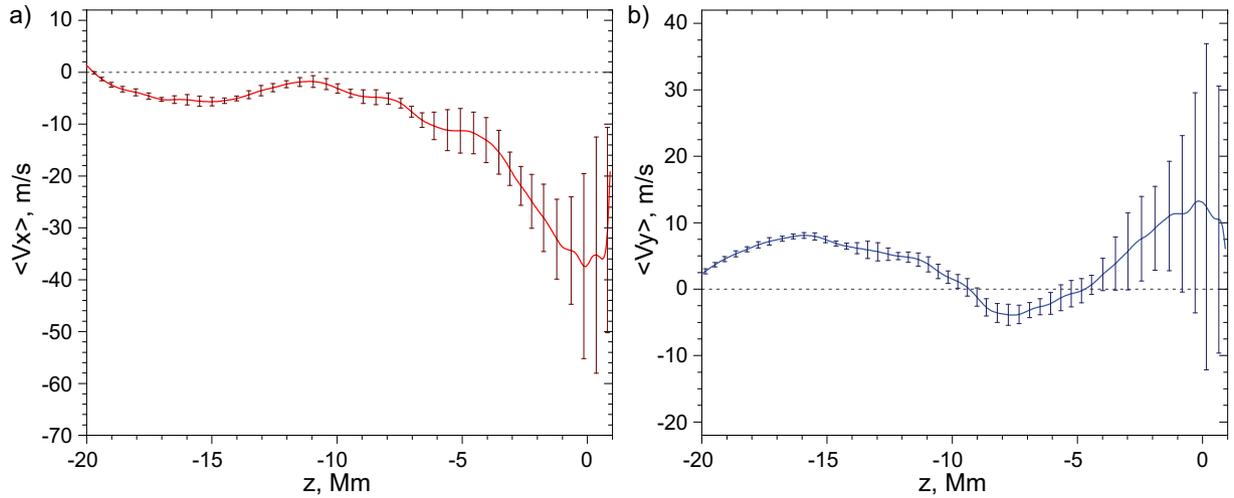}
	\end{center}
	\caption{Mean radial profiles of a) deviations of the azimuthal flow speed from the imposed rotation rate at 30 degrees latitude, b) the meridional component of the flow velocity. The vertical bars show $1\sigma$ flow velocity deviations from the mean values. Radial profiles are obtained by averaging a 24-hour series of simulation data. \label{fig:Vxyz2}}
\end{figure}

To inspect deviations of the azimuthal flows from the imposed rotation, we calculate the radial profile of velocity along the direction of rotation ($<Vx>$) averaged in the horizontal directions and time. The results, shown in Figure~\ref{fig:Vxyz2}a, reveal a significant decrease in the azimuthal velocity with depth by 38~m/s in a 2~Mm deep layer below the photosphere. Below 7~Mm, the rotation rate is slower than the imposed mean rotation rate by about 5~m/s. Interestingly, the rotation rate increase with depth is not uniform: from the sub-photospheric layers to about 4~Mm below, the velocity increases by $6 - 7$~m/s per Mm, while below 4~Mm the flow accelerates by about 2~m/s per Mm.
A similar change in the differential solar rotation rate at similar depths has been demonstrated in pioneering helioseismology observations by \cite{Deubner1979} in their Figure~\ref{fig:obs}, where analysis of the $k-\omega$ diagram of solar oscillations showed a noticeable increase of the relative horizontal flows.
The identified 10-Mm thick near-surface shear layer, or `leptocline' \cite[named by analogy from the tachocline and originated from greek {\it leptos} that means `fine',][]{Godier2001,Rozelot2009b}, is clearly visible in the relative differential rotation profile (Figure~\ref{fig:Vxyz2}a). 

The meridional component of the mean velocity ($<Vy>$, Figure~\ref{fig:Vxyz2}b) reveals a complex structure with mostly poleward flows and a speed of $\sim 12-13$~m/s in the near-surface layers and about 8 m/s at a depth of 16 Mm. The meridional component of flow decelerates from the photosphere from 12~m/s to about --4~m/s at a depth of 8~Mm. Thus, a weak reverse flow occurs at $5 - 10$ Mm depth. Below 8 Mm, the meridional flows accelerate again in the poleward direction.

\begin{figure}
	\begin{center}
		\includegraphics[width=0.9\textwidth]{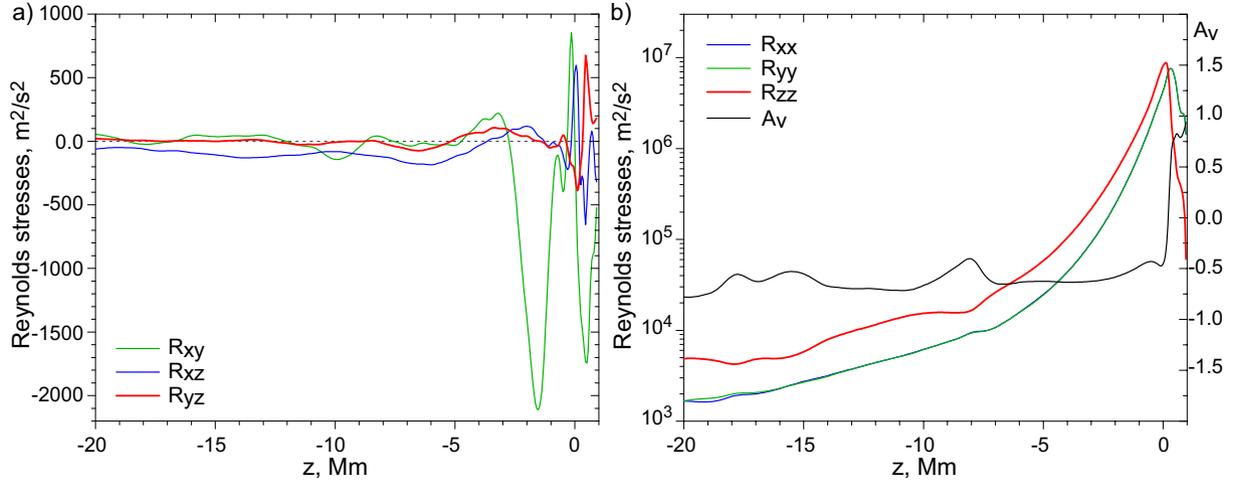}
	\end{center}
	\caption{Mean radial profiles of a) off-diagonal Reynolds stresses, and b) diagonal Reynolds stresses and the anisotropy parameter, $A_V$ (black curve). Radial profiles are obtained by averaging a 24-hour series of simulation data. \label{fig:Rey}}
\end{figure}

The distribution of Reynolds stresses (computed as $R_{ij}=<u'_iu'_j>$, where $u'_i$ and $u'_j$ are the velocity component fluctuations) reveals a complex coupling between the large-scale flows and small-scale turbulent motions (Figure~\ref{fig:Rey}a). It is not surprising that variations of the Reynolds stresses are strongest near the photosphere.
In the absence of rotation or if rotation is too slow to influence the turbulence, it is expected that the mean horizontal component of the Reynolds stresses, $R_{xy}$, has minimal variations. 
However, as shown in Figure~\ref{fig:Rey}a, the horizontal Reynolds stresses vary significantly from the low atmosphere down to layers about 4 Mm deep. In particular, strong variations of $R_{xy}$ with a peak of $-2111$~m$^2$/s$^2$ at a depth of 1.5~Mm correlate with the bottom of the granulation layer. The longitudinal (or azimuthal) component of the Reynolds stresses  ($R_{yz}$, red curve) reveals strong variations near the photosphere. Below the photosphere,  $R_{yz}$ variations are weaker and vary around zero below 5~Mm.
The meridional component of the Reynolds stresses is negative at the photosphere, revealing a sign change at the near-surface layers, where it reaches a maximum of $\sim 100$ m$^2/$s$^2$ at a depth of 2~Mm. Below that, $R_{xz}$ gradually decreases down to about 6~Mm below the surface and then fluctuates around $-90$~m$^2$/s$^2$ in deeper layers of the convection zone. 

In layers deeper than 5~Mm, there appears to be no preferred sign for horizontal Reynolds stresses {\bf $R_{xy}$}, except in a 1 -- 2~Mm thick layer at a depth of 10~Mm that indicates the presence of horizontal shear (green curve, Figure~\ref{fig:Rey}a). We identify this layer as the bottom of the leptocline. This interface between the leptocline and deeper layers of the convection zone is manifested as a `kink' in the horizontal diagonal Reynolds stress components and a `pit' in vertical one (Figure~\ref{fig:Rey}b), which signifies  overshooting downdrafts from the highly convectively unstable hydrogen ionization layers into a less unstable layer between the H and HeII ionization zones. Similar changes in radial profiles of the horizontal and vertical flows were found in our previous  simulations of convective overshooting at the bottom of the convection zone of a more massive star \citep{Kitiashvili2016}. The bottom boundary of leptocline is also pronounced as a peak in the anisotropy of the horizontal and vertical turbulent flows $ A_V=(R_{xx}+R_{yy}-2R_{zz})/V^2_{rms}$ (black curve, Fig.~\ref{fig:Rey}b). This feature was not found in the recent simulations of the NSSL by \cite{Barekat2021}, in which effects of stratification were not included. 

The distribution of the off-diagonal stresses is important for the interpretation of differential rotation and meridional circulation in terms of the mean-field theory \citep{Ruediger1989}. The detailed mean-field analysis is outside the scope of this paper, and here we make only some qualitative remarks. According to this theory, the angular momentum in the anelastic approximation is established through the balance of the Reynolds stresses, $R_{xy}$, and the transport by meridional flows \citep[e.g., Eq.~7.27 in  ][]{Stix2002}. The values of $R_{xy}$ are negative in the leptocline and become positive near the bottom of the leptocline; this is in the qualitative agreement with the direction of the meridional circulation in our simulations. The radial gradient of the rotational velocity is determined by a balance of diffusive and non-diffusive (so-called Lambda-effect) components of the off-diagonal Reynolds stresses \citep[Eqs.~31-32 in ][]{Barekat2021}. Because $R_{xz}$ is predominantly negative (Fig.~\ref{fig:Rey}a), the vertical Lambda-effect coefficient must be negative. This is consistent with the results of \cite{Barekat2021}. However, contrary to their simulations, we find that the horizontal Reynolds stresses, $R_{xy}$, are dominant in the leptocline. Thus, further studies are needed to obtain a consistent picture of the turbulent Reynolds stresses in the near-surface shear layer, and to understand their effects on the large-scale dynamics.

\begin{figure}
	\begin{center}
		\includegraphics[width=0.9\textwidth]{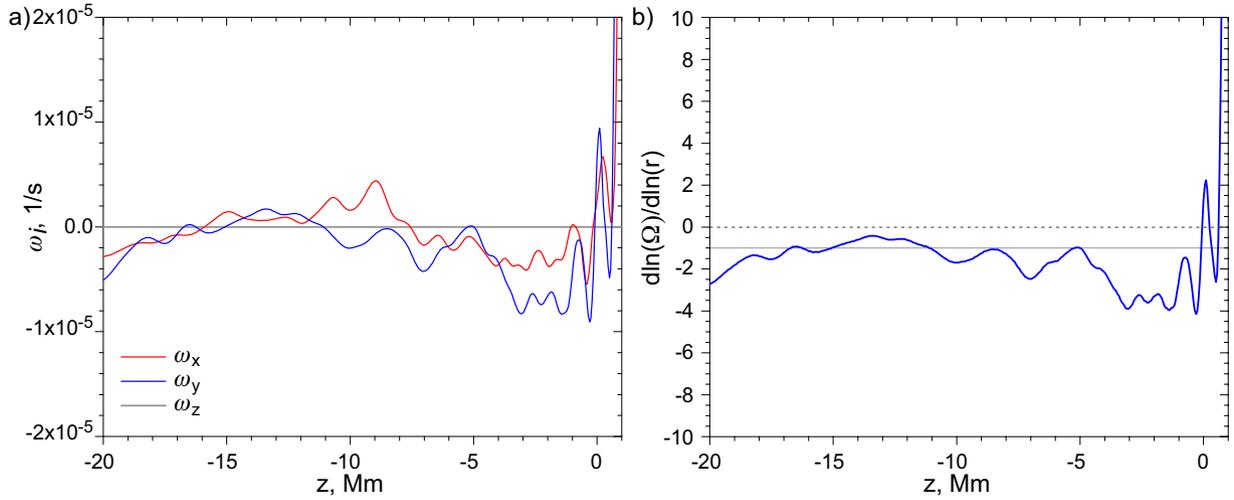}
	\end{center}
	\caption{Mean radial profiles of a) the vorticity components, and b) the radial gradient of the local solar rotation rate, defined as  $\frac{\partial\ln\Omega}{\partial\ln r}$. Radial profiles are obtained by averaging a 24-hour series of simulation data. \label{fig:vort}}
\end{figure}

In the presence of rotation, the radial profile of the mean vertical vorticity distribution ($\omega_z$, Figure~\ref{fig:vort}a, gray line) does not indicate preference of the vortical motions. On the other hand, the horizontal vorticity components (blue and red curves) reveal negative values, mostly in the top 5-Mm of the subsurface layers, which indicates a preference for clockwise rotational turbulent flows. There is no significant directional preference for horizontal vorticity in the deeper layers, where the turbulence becomes more isotropic. A decrease of the horizontal vorticity around 18 -- 20~Mm below the surface potentially due to the closed bottom boundary for flows at depth $-25$Mm and requires additional investigation. 

Helioseismic measurements {have shown} that the radial gradient of solar rotation, 
$\frac{\partial\ln\Omega}{\partial\ln r}$, 
(where $\Omega$ is the local angular velocity, $r$ is the radius) varies with latitude \citep{Corbard2002}, and has a value of about $\sim -1$ from the equator to 30~degrees latitude in the outer 15~Mm layer of the convection zone. At higher latitudes, the gradient is negative but has smaller values. These results have been confirmed by \cite{Barekat2014}.
More recent studies showed more substantial variations of the radial gradient with latitude. Also, the inferred radial profile depends on the selected range of spherical degrees of solar oscillation modes used in the inversion procedure \citep[e.g., $\sim-2.8$ for the high-degree inversions and $-2.13$ for the intermediate-degree inversion at 30 degrees latitude;][]{Reiter2020}. According to recent helioseismic studies by \cite{Antia2022}, the gradient changes with depth from $\sim -0.95$ at a depth of 7~Mm to $-0.55$ at a depth of 20~Mm.

Our results cover layers from the photosphere to 20 Mm below and show stronger negative values of the gradient of rotation, about $-4$ in subsurface layers, and an increase in the deeper layers Figure~\ref{fig:vort}b). Interestingly, the rotation gradient shows qualitatively the same variations as the meridional component of vorticity, $\omega_y$, (blue curve, Figure~\ref{fig:vort}a), which indicates coupling of the large-scale flows and turbulence. 
In the model, the gradient of rotation is shifted to lower values by about unity from the photospheric value obtained from high-spherical degree helioseismic inversions by \cite{Reiter2020}, and values obtained by \cite{Antia2022} for depths of 7~Mm and 20~Mm. Taking into account that the helioseismology inferences have significant averaging over depth (with ``averaging kernels''), and do not provide a good localization of the averaging kernels near the surface, we can conclude that the modeled radial profile of the radial gradient of the rotation rate is in reasonable agreement with the observations. 

\begin{figure}
	\begin{center}
		\includegraphics[width=0.9\textwidth]{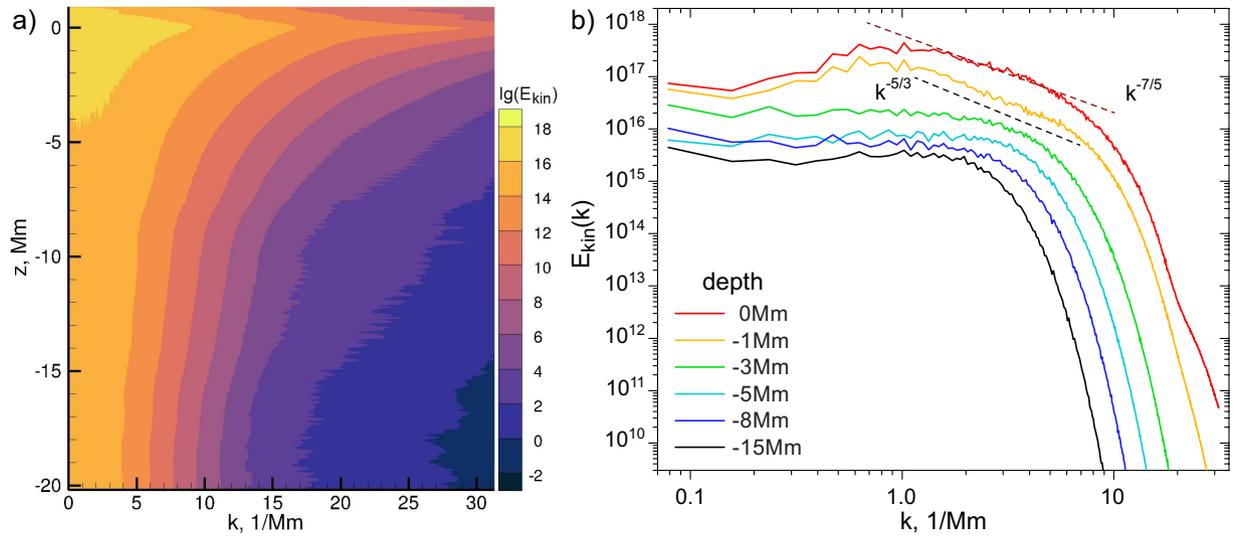}
	\end{center} 
	\caption{Averaged velocity power spectra. Panel a: Distribution with depth. Panel b: Spectra for several selected depths: from 15 Mm (black curve) to the photosphere (red). \label{fig:turb}}
\end{figure}

Because of the complexity of the mean flow distribution, it is interesting to consider how the velocity power spectra change with depth in the convection zone (Figure~\ref{fig:turb}). As expected, the power is mainly concentrated in the near-surface flows and gradually decreases with depth. The rate of decrease reveals two sublayers (Fig.~\ref{fig:turb}a): 1) subsurface layers up to 7 -- 8~Mm, with a fast decrease, and 2) below 8~Mm, with a slow decrease. A spectral slope of $k^{-5/3}$ corresponds to a Kolmogorov-type inertial range at 1 Mm depth (Fig.~\ref{fig:turb}b). Similar changes with depth of the turbulent properties were previously demonstrated in simulations for a small (6.4~Mm wide and 5~Mm deep) computational domain \citep{Kitiashvili2013}.  
	A significant reduction in level of the velocity power distribution and the disappearance of the inertial range in the spectra for the deeper layers likely reflects a decrease of the  Rayleigh number with depth. A weak increase of the kinetic energy at small wavenumbers, $\sim 0.2$~Mm$^{-1}$, indicates the presence of a convection scale of $\sim$20~Mm that is comparable with the supergranulation scales. This scale becomes more prominent in the deeper layers. A more detailed study of this scale is required using a larger and deeper computational domain with a background magnetic field to determine its relation, if any, to observed supergranulation properties.

\begin{figure}
	\begin{center}
		\includegraphics[width=0.9\textwidth]{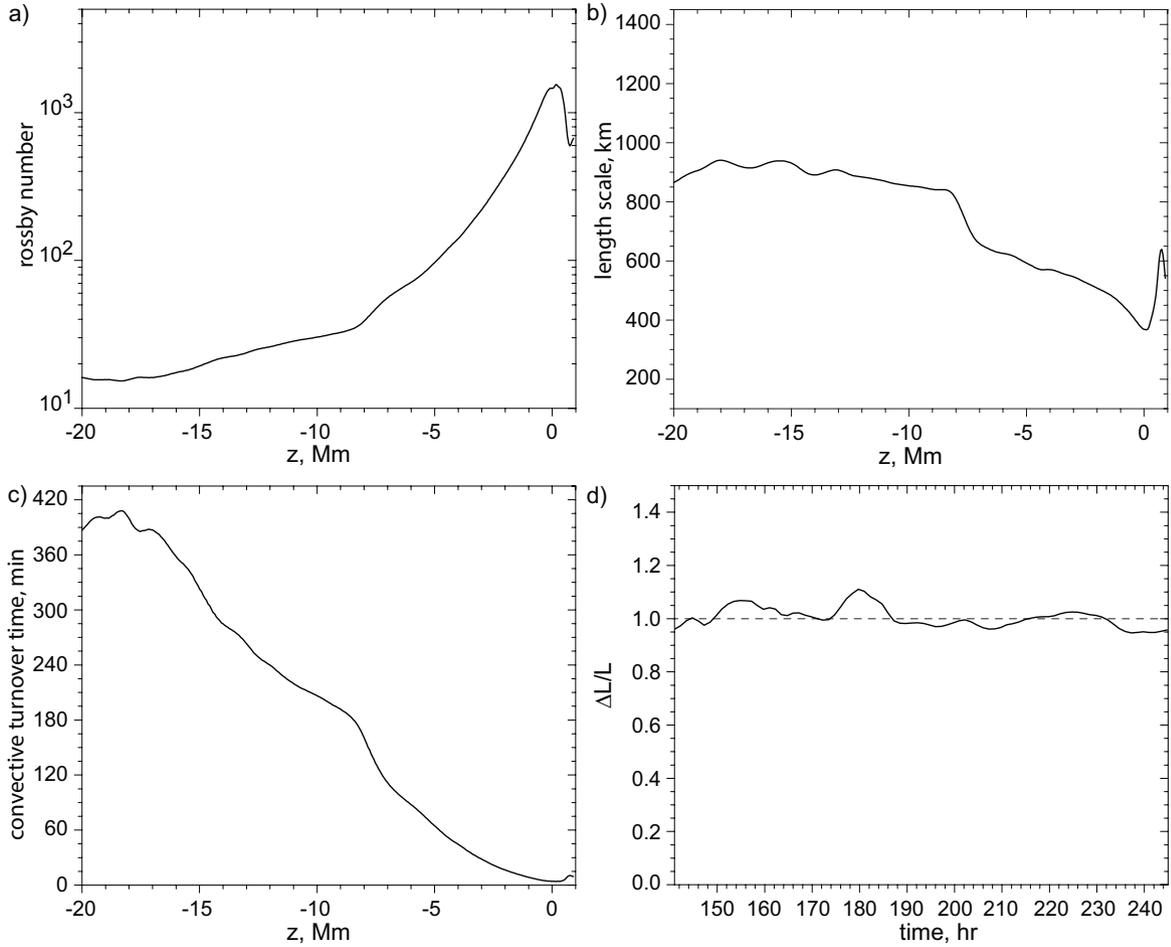}
	\end{center}
	\caption{Radial profiles of the Rossby  number (panel a), characteristic length scale (panel b), convective turnover time (panel c), and time-evolution of the relative angular momentum (panel d). Radial profiles on panels a-c) are time-averaged over 1 hour. \label{fig:rossby}}
\end{figure}

In the presence of rotation, it is natural to consider convection zone properties in terms of the Rossby number, the length scale of the turbulence, and the convective turnover time (Figure~\ref{fig:rossby}). Following \cite{Guerrero2019}'s suggestion, the length scale of the turbulent plasma is calculated as 
\begin{displaymath}
\ell(r)=\frac{r\int_k \frac{\tilde{E}(k,r)}{k} dk}{\int_k \tilde{E}(k,r) dk}, 
\end{displaymath}
where $\tilde{E}(k,r)$ is the velocity power spectral density, $r$ is the solar radius, and $k$ is the wavenumber. The convective turnover time is then $\tau_c=\ell(r)/V_{rms}$. The Rossby number can be expressed as $Ro=P_{rot}/(2\pi \tau_c)$, where $P_{rot}$ is the rotational period.

According to our model, the Rossby number is the highest at the solar photosphere, where the turbulent flows are the strongest, and the convective turnover time is the shortest (Figure~\ref{fig:rossby}). It is known that the turbulent length-scale and the convective turnover time gradually increase with depth, resulting in a gradual decrease of the Rossby number in deeper layers. Interestingly, the length scale below the hydrogen ionization zone is almost constant, with variations only around 100~km in the layers from 8 to 20~Mm below the solar surface (Figure~\ref{fig:rossby}b). However, the Rossby number decrease is not uniform. In particular, near the bottom of the hydrogen ionization zone, at a depth of 7 Mm below the surface, the length scale and the turnover time suddenly increase, thus slowing down the decrease of the Rossby number.

Because of the density stratification, it is natural to consider how the temperature and density perturbations vary with radius (Figure~\ref{fig:TRhoRmsGK}a). In particular, the RMS temperature fluctuations (red curve) have a strong peak at the solar photosphere and exponentially decrease with depth. The radial profile becomes steeper at a depth of 8~Mm with a sudden reduction in temperature fluctuations. The thickness of the layer is about 1~Mm. The density fluctuations (blue curve) also reveal a sharp increase in the photosphere. In general, the RMS density fluctuations increase with depth up to 5~Mm below the photosphere and then become saturated. Similar to the temperature variations (red curve), the density variations sharply decrease in a 1 Mm-thick layer, near a depth of 8 Mm. This layer is located at the bottom of the hydrogen ionization zone (Figure~\ref{fig:TRhoRmsGK}b). The adiabatic exponent $\Gamma_1$ has a bump between 8 and 10 Mm, resulting in a layer of weaker convective instability between the hydrogen and second-helium ionization zones. This leads to convective overshooting effects, probably related to formation of the shallow return meridional flow. Curiously, the shallow convective layer in the hydrogen ionization zone (the leptocline) seems to have some properties analogous to ones for the whole convection zone, such as convective overshooting, a tachocline, and return meridional flow. 

\section{Discussion and Conclusions}\label{conclusion}
To investigate effects of solar rotation on the dynamics and structure of the upper 20 Mm of the solar convection zone, we performed 3D radiative hydrodynamics simulations for 250 hours of solar time that allowed us to achieve quasi-stationary dynamical conditions and provided a long dataset of more than 100 hours for analysis. 
The simulation results reveal the development of radial differential rotation (Fig.~\ref{termdyn}a) with the strongest gradient in a 10 Mm-thick subsurface layer (a so-called leptocline).
The leptocline represents an upper layer of the well-known Near-Surface Shear Layer (NSSL) that extends up 35-Mm deep (about 5\% of the solar radius). Because the whole computational domain of our model is 25-Mm deep, we do not model the dynamics and structure of the whole NSSL and, in particular, its interaction with the deep convection zone because it requires a significantly deeper computational domain and more computational resources. However, we focus on studying the structure and dynamics of the upper part of the NSSL. We show that the outer layers of the NSSL form a distinct substructure characterized by enhanced turbulent convection and strong rotational shear, which we identify as leptocline, a shallow subsurface layer conjectured in previous publications  by \cite{Godier2001} and \cite{Rozelot2009b}. Similar to the NSSL, it reveals the properties of the shear layer. The interaction of the leptocline with deeper layers of the NSSL is caused by overshooting of convective downdrafts formed in the leptocline into the relatively less convectively unstable layers between the H and He\,II ionization zone. This overshooting layer defines the bottom boundary of the leptocline and makes it dynamically distinct from the deeper NSSL. 
A signature of the leptocline corresponding to a similar depth range first was detected by \cite{Deubner1979} and in most recent helioseismic studies \citep[e.g.,][]{Reiter2020,Komm2021,Antia2022}.

\begin{figure}
	\begin{center}
		\includegraphics[width=0.45\textwidth]{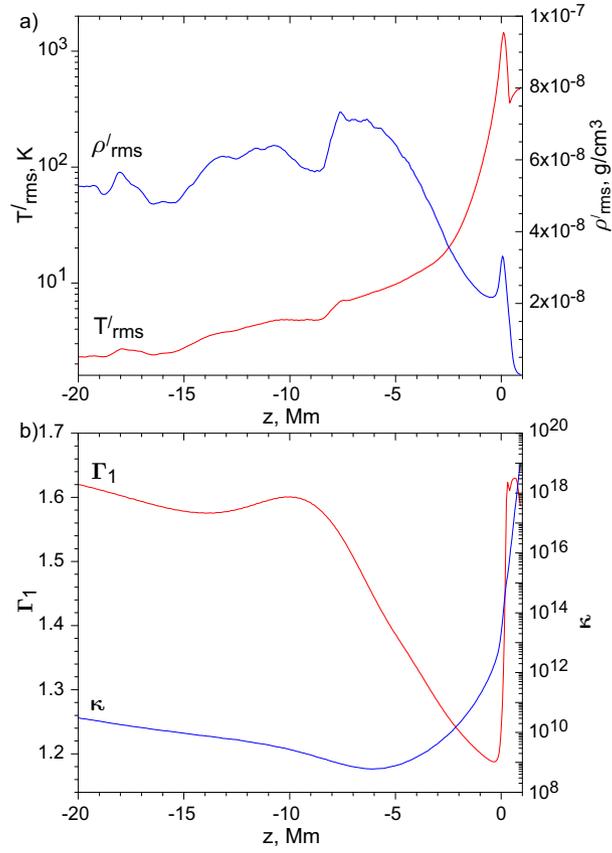}
	\end{center}
	\caption{Panel a: Radial profiles of the temperature perturbations, $T'_{rms}$ (red curve), and density perturbations, $\rho'_{rms}$ (blue curve). Panel b: Radial profiles of the adiabatic index, $\Gamma_1$ (red curve) and the heat conductivity (blue curve). Radial profiles are time-averaged over 24 hours.} \label{fig:TRhoRmsGK}
\end{figure}

\begin{figure}
		\begin{center}
	\includegraphics[width=0.45\textwidth]{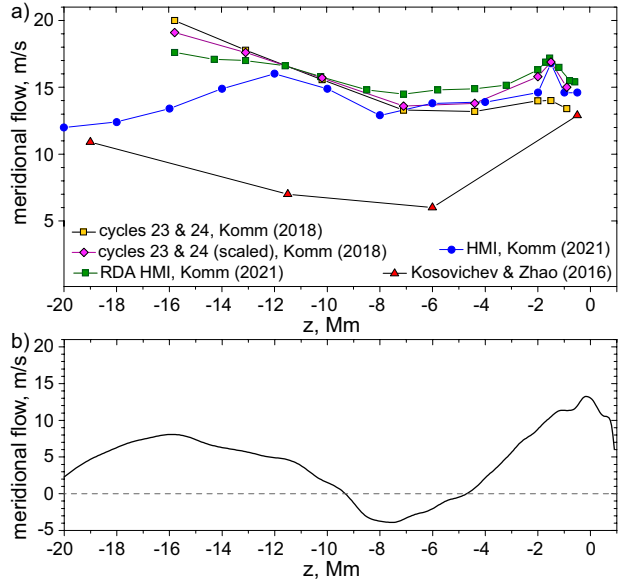}
		\end{center}
	\caption{Panel a) Meridional circulation helioseismic inferences at 30 degrees latitude. Panel b shows qualitative agreement of the mean profile of the meridional component of flows with observations (panel a).}
	\label{fig:obs}
\end{figure}

This boundary layer is characterized by an accelerated reduction of the Rossy number with depth and an increase in the length scale and convective turnover time  (Fig.~\ref{fig:rossby}). In terms of the thermodynamic properties, the bottom boundary of the leptocline is associated with a noticeable decrease of the temperature and density fluctuations (Fig.~\ref{fig:TRhoRmsGK}a) and is related to the bottom of the hydrogen ionization zone (Fig.~\ref{fig:TRhoRmsGK}b). The sound-speed profile in the helioseismic inversions of high-degree solar oscillation modes reveals a sharp change at a 5-6 Mm depth \citep[see Figure~B4 in][]{Reiter2020}, which {is probably} be related to the existence of the leptocline.

The resulting rotational velocity reduction relative to the imposed rotation rate of $\sim40$~m/s at the photosphere is in agreement with observations near the solar minimum obtained with a variety of techniques  \citep[e.g.][]{Imada2020}. It is important to note that the presence of a leptocline can be identified in multiple observations. The first indication of a 10-Mm thick subsurface layer was demonstrated by \cite{Deubner1979}, using three one-day-long data series of observations, where inferences were made for every 2-Mm depth range up to 20~Mm below the photosphere. 
The recently developed new methods to perform more accurate rotational inversions allow distinguishing the leptocline in the radial rotational profiles \citep[see Figure~27 in ][]{Reiter2020}. According to these results, the leptocline is manifested as a steeper slope of the rotation rate near the photosphere, followed by a well-known subsurface shear layer with a weaker angular velocity gradient. Another helioseismic investigation, based on the ring-diagram analysis of SDO/HMI and GONG Dopplergrams, shows a qualitative change in the properties of the zonal flows at a depth of 8--10-Mm \cite[see Figure 3 in ][]{Komm2021}, where the leptocline signature becomes more prominent for higher latitudes.

Another interesting feature of the leptocline is the presence of a convective overshoot at the bottom of the hydrogen ionization zone that intensifies the mixing of the turbulent flows. The overshoot layer (interface between the leptocline and deeper layers of the convection zone), was revealed from an analysis of the Reynolds stresses and variations of the velocity magnitude (Figs~\ref{fig:Rey}a and~\ref{fig:rossby}d) that indicate a splashing of the downflows. This interpretation is also supported by significant variations of the temperature and density fluctuations at the bottom of the hydrogen ionization zone  (Fig.~\ref{fig:TRhoRmsGK}). The interface between the leptocline and deeper NSSL is well revealed as a peak in the anizotropy of Reynolds stresses (parameter $ A_V$ in Fig.~\ref{fig:Rey}b).

In addition, our model with imposed rotation reveals the formation of meridional flow. At the photosphere, the flow is in the northward direction with a mean speed of 12 -- 13~m/s. At a depth of 4~Mm, the flow changes direction to equatorward and reaches $\sim -4$~m/s at a depth of 8~Mm. 
The reverse meridional flows correspond to the interface between the shallow sub-surface shear layer discussed above (leptocline) and the rest of the convection zone. In the deeper layers, the flows accelerate again and change direction to northward at a depth of 10~Mm. At a depth of 16~Mm, the mean velocity reaches the speed of  8~m/s. 
The described sub-surface flows suggest a possibility of fine structuring, similar to previously discovered two-cell meridional circulation on the scale of the whole convection zone \citep{Zhao2013,Chen2017}.

It is known that the properties of the meridional flows vary during the solar cycle. For comparison with observations, we consider only observational studies that occur at a minimum of solar activity (or close to it) at 30 degrees latitude as the most relevant conditions to the hydrodynamic simulations. In general, we found a good agreement of the resulting meridional flows with surface \citep[e.g.,][]{Ulrich2010,Imada2018} and subsurface helioseismic inferences \citep[e.g.,][]{Zhao2012,Komm2018,Antia2022}. 
The previous helioseismic studies reveal a qualitatively similar distribution of the meridional flows \citep[e.g.][]{Kosovichev2016,Komm2018}: deceleration from the photosphere to 8--10 Mm and its acceleration at the deeper layers. In particular, Figure 8b in \cite{Kosovichev2016} shows that the meridional flows are getting weaker at a depth of 6 -- 11.5~Mm and become stronger again with values comparable with the near-surface speed at a depth of 19~Mm. 
Figure~\ref{fig:obs}a summarizes some helioseismic inferences obtained with the ring-diagram analysis \citep{Komm2018,Komm2021} and the time-distance method \citep{Kosovichev2016} in the NSSL. Panel b) is given for the reference to the modeled meridional flows. The previously published helioseismic inferences are obtained for different time intervals and show diverse values in the meridional flow speed. Still, all of them reveal deceleration of the flows in layers corresponding to the bottom of the leptocline. Extended flow measurements by \cite{Komm2021} show a secondary deceleration of flows for layers deeper than 12 Mm. Discrepancies between observations and simulation can be explained by three factors: 1) the model is obtained in the local box, therefore, cannot take into account the `thermal wind' and other global-Sun effects \citep[e.g.,][]{Spiegel1992}, 2) helioseismic inferences are obtained with significant depth averaging \citep[e.g., the depth extent of averaging kernels in time-distance inversions exceeds 4 Mm;][]{Couvidat2005}, 3) presence of the solar activity affects meridional flows \citep[e.g.,][]{Getling2021,Komm2021}. Thus, further modeling and helioseismic studies are required for determining the structure  of the meridional flows in the NSSL.	
Therefore, the resolution of the helioseismic inversions was not sufficient to resolve the shallow layer with the return flow; however, it was observed as a reduction in the flow speed. Indeed, more precise helioseismic measurements of the leptocline are needed. In addition, high-resolution global-scale modeling is required to understand the role of the leptocline in global-Sun dynamics and activity.

To summarize the presented analysis of the 3D radiative hydrodynamics simulations with imposed rotation corresponding to 30 degrees latitude, we can identify the following main results:
\begin{itemize}
	\item The simulations reveal the development of radial differential rotation and formation of a shallow $\sim$10~Mm-deep substructure of the Near-Surface Shear Layer -- the leptocline -- associated with a steep radial gradient of the angular velocity, changes in the thermodynamic and structural properties of the convection, and the bottom of the hydrogen ionization zone. 
	\item The leptocline constitutes the upper part of the Near-Surface Shear Layer. The interface between the leptocline and deeper layers is characterized by overshooting downdrafts, which may intensify the turbulent mixing below the leptocline layer.
	\item  Our results show that the Near-Surface Shear Layer is not uniform with a constant rotational gradient, but it has an important shallow substructure -- the leptocline.
	\item The self-formed meridional flows are characterized by poleward mean motions near the photosphere and weak reverse flows at depths of $5 - 10$~Mm. The bottom of the reverse flows corresponds to the bottom of the leptocline layer and the hydrogen ionization zone. This structure resembles a double-cell meridional circulation previously found on the whole-convection-zone scale.
	\item Analysis of the turbulent spectra indicates the presence of convective patterns with the scale of $\sim$~20 Mm that closely corresponds to the scale of supergranulation.
\end{itemize}

The next step of this work is to perform modeling for different latitudes and investigate the latitudinal structure and dynamics of the leptocline.

\section*{Data Availability Statement}
Results of the analysis data underlying this article will be shared on request to the corresponding author. \label{key}

\section*{Acknowledgements}
The modeling and data analysis of the resulting data are performed using the NASA Ames Supercomputing Facility. The presented investigation is supported by NASA Heliophysics Supporting Research Program, NASA grant: NNX14AB70G, and NASA DRIVE Science Centers grant: 80NSSC20K0602 (COFFIES).



\bibliographystyle{mnras}


\onecolumn

\bsp	
\label{lastpage}
\end{document}